# Aluminium substitution induced superstructures in $Mg_{1-x}Al_xB_2$ (x = 0.0 to 0.50): An X-ray diffraction study

Monika Mudgel[1], V.P. S. Awana[1, *], H. Kishan[1] and G. L. Bhalla[2]

[1]National Physical Laboratory, Dr. K.S. Krishnan Marg, New Delhi-110012, India

[2]Deptartment of Physics and Astrophysics, Delhi University, New Delhi, India

## Abstract

The physical property characterization of Al doped $Mg_{1-x}Al_xB_2$ system with x = 0.0 to 0.50 is reported. The results related to phase formation, structural transition, resistivity $\rho(T)$ and magnetization $M(T)$ measurements are discussed in detail. It is shown that the addition of electrons to $MgB_2$ through Al results in loss of superconductivity. Also seen is a structural transition associated with the collapse of boron layers reflected by the continuous decrease in the *c* parameter. The main emphasis in this paper is on slow scan *X*-ray diffraction (*XRD*) results, which confirm the existence of a superstructure along the *c*-direction for the x = 0.50 sample. The appearance of some additional peaks, viz. [103], [004], [104] and [112], results in doubling of the lattice parameter along the *c*-axis. This possibly indicates the alternative ordering of Al and Mg in $MgAlB_4$ separated by hexagonal boron layers but still maintaining the same hexagonal $AlB_2$ type structure.

*: Corresponding Author; e-mail-awana@mail.nplindia.ernet.in

## Introduction

After the announcement of superconductivity in $MgB_2$ [1], everyone hoped that this material would be the starting one in the series of superconducting diborides. The simple hexagonal structure and composition attracted the attention of many groups towards the study $MgB_2$ [2, 3] and similarly structured diborides [4-6]. But still $MgB_2$ holds the record of the highest $T_c$ among all diborides, like $TaB_2$, $MoB_2$, $ZrB_2$, etc. Interestingly, $AlB_2$ is not a superconductor at all, which raises curiosity about the mechanism of superconductivity in diborides other than $MgB_2$. There is evidence for the well-established phonon mediated BCS superconductivity in $MgB_2$ [7,8], and it's high $T_c$ (39 K) value is believed to be due to the high phonon frequencies and strong electron phonon interactions. This theory is supported by the results of various experiments, such as isotope effect [2,9] and specific heat measurements [10]. A universal mechanism for superconductivity in $MgB_2$ concerning the pairing of *dressed* holes was put forward by Hirsch [11]. Indeed the hole character of the carriers in $MgB_2$ was confirmed by Hall [12] and thermoelectric power measurements [13,14]. But in order to understand the superconducting properties in similarly structured diborides, partial chemical substitution in $MgB_2$ can be regarded as a good method. So researchers started looking at the substitution chemistry of $MgB_2$. Although substitution of other elements into $MgB_2$ is not an easy task, Al appears to be an exception since it has comparable ionic size with Mg. Al substitution has been extensively studied both experimentally and theoretically but for low x values [15-18]. Various studies, like Raman spectroscopy, infrared spectroscopy, electron energy loss spectroscopy, optical spectroscopy, band calculations, heat capacity measurements, etc. have been done on this system (19-23). The $c/a$ value is 1.143 for $MgB_2$ while it is 1.083 for $AlB_2$, which clearly indicates that $MgB_2$ is stretched in the $c$ direction in comparison to $AlB_2$. Worth mentioning is the fact that there is not much difference in the ionic sizes of Mg and Al. Hence, the stretching of the $MgB_2$ lattice in comparison to $AlB_2$ can not be explained simply on the basis of crystal chemistry and electronic changes must be probed. Keeping all this in mind, we substituted Al into the $MgB_2$ lattice in various amounts and analyzed the structural and superconducting properties of the doped system. Al substitutes into Mg successfully at low concentration and single-phase purity is observed. But at intermediate concentrations some complexities arise in terms of unusual peaks in the *XRD* patterns, which are analyzed in terms of a superstructure. This superstructure is related to the structural transition in terms of Mg and Al ordering in the $MgAlB_4$ lattice. The superconducting transition temperature ($T_c$) decreases with x in the $Mg_{1-x}Al_xB_2$ system. In the current article, the Al induced structural changes in the $Mg_{1-x}Al_xB_2$ system ensuing superstructure at x = 0.50 are discussed, mainly based on results from room temperature *X*-ray diffractometry. The superconducting properties of the system are also discussed briefly.

## Experimental

Polycrystalline samples of $Mg_{1-x}Al_xB_2$ (x = 0.0 to x = 0.50) were synthesized by the conventional solid-state reaction route. High purity Mg, B, and Al powders prepared in stoichiometric ratios were thoroughly ground for two hours. Then, the homogenous mixtures so obtained were pelletized using a hydraulic press by applying a pressure of 7.5 tons/cm$^2$. These pellets were enclosed in open-end safety Fe-tube and subsequently heated at a temperature of 850$^o$ C in an Ar atmosphere at ambient pressure. The temperature was ramped at a rate of about 425$^o$ C per hour followed by a holding time of two and a half hours. The samples were slow cooled in the same atmosphere down to room temperature. The samples obtained were characterized structurally, electrically, and magnetically. *X*- ray diffraction patterns were taken using Ni Filtered $CuK_\alpha$ radiation at 40 kV and 50 mA with a Rigaku RINT2200HF-Ultima diffractometer. The obtained samples were hard and dense enough for resistivity measurements. Resistivity measurements were made on bar shaped samples using the four-probe technique. Magnetic susceptibility measurements were also performed to determine the superconducting transition temperature on a Quantum designed SQUID magnetometer (*MPMS-XL*). In the zero field cooled case, samples were first cooled in zero field down to a temperature below their critical temperature. Subsequently, a low field was applied and the diamagnetic moment as a function of slowly increasing temperature was recorded, while in the field cooled case the samples were cooled in the presence of a small field of about H = 10 Oe and magnetization measurements were done while cooling.

## Results and Discussions

The X- ray diffraction pattern for pure $MgB_2$ is shown in Figure 1(a). All characteristic peaks are obtained which are in good confirmation with the literature [1-3, 12-14]. Indexing of all peaks is marked on the pattern. A low intensity peak at $2\theta = 62.5\,^o$ indicates the presence of a minor amount of MgO as an impurity phase [24, 25]. X-ray diffraction patterns for $Mg_{1-x}Al_xB_2$ samples (x = 0.0 to x = 0.40) at room temperature are shown in Fig.1 (b). All samples crystallize in the simple hexagonal $AlB_2$ type structure with space group *P*6/*mmm*. The lattice parameters for all samples are calculated and tabulated in Table 1 which matches well with the reported literature [10, 16]. There is a small change in the *a* parameter as Al content increases. It can also be seen from the upper inset in the *XRD* pattern shown in Fig. 1(b) that there is a shift in the [100] peak towards the higher angle side showing a decrease in the *a* parameter. On the other hand there is a relatively large change in the *c* parameter due to a considerable shift in the [002] peak, again towards the higher angle side, shown clearly in the lower inset of Fig. 1(b). It results in a decrease of the *c* parameter that is about three times larger than the decrease in the *a* parameter, which means that Al substitution does not have much effect on the intra planar distance between the boron atoms, i.e., it confirms the rigidity of the boron honeycomb layer as reported by Slusky et al [15]. But the much pronounced decrease in the *c* parameter in turn confirms the partial collapse between the boron layers. In this way a negative strain (reflected by continuous decrease in both lattice parameters and hence a net decrease in cell volume) is introduced in the $MgB_2$ lattice on addition of Al. Along with the shift, broadening of (002) peak is also observed. All this structural information is supported by various other groups [15, 26, 27-29] confirming the good quality of our samples. Interestingly enough, at higher concentration (> x = 0.20) some extra peaks are observed (marked with #, in Fig. 1(b)) for say x = 0.40 in the *XRD* pattern.

Fig.2 depicts the exact variation of the *a* and *c* lattice parameters of the $Mg_{1-x}Al_xB_2$ system. Both *c* and *a* decrease with increasing Al content. The *c*/*a* value is also plotted and shown in the inset of Fig. 2. A continuous decrease is observed in *c*/*a* value with increasing Al doping. This in turn shows that the decrease in the *c* parameter is quite large compared to the decrease in the *a* parameter.

Fig. 3 depicts the XRD pattern of the $Mg_{0.5}Al_{0.5}B_2$ sample. The extra peaks observed at x = 0.40 in the range $2\theta = 50^o$ to $55^o$ and $2\theta = 62^o$ to $68^o$ are found to be more dominant in the $Mg_{0.5}Al_{0.5}B_2$ sample (see Fig. 3). These extra peaks cannot be indexed by a single cell approach even with decreasing *a* and *c* lattice parameters. But we can successfully index these extra lines as [103], [004] and [104], [112] in the vicinity of [002] and [102], respectively, by taking into account the concept of a double cell along the *c* axis. This can be seen clearly in the left and right hand side insets of Fig. 3. The cell refinement report for both $MgB_2$ and $Mg_{0.5}Al_{0.5}B_2$ is shown in Table 2. The observed $2\theta$ values of peaks match well with the calculated $2\theta$ values in both cases. All theoretically determined peaks for $MgB_2$ are observed in the pattern. For $Mg_{0.5}Al_{0.5}B_2$, two peaks namely (001) and (003). are missing. The double cell concept is taken into account for $Mg_{0.5}Al_{0.5}B_2$. In figure 3, the symbol $ represents the undefined peaks and the # sign indicates the peaks corresponding to the superstructure. Two more peaks having low intensities are also seen on both sides of the main peak (102) in this sample, The left-hand peak corresponds to $MgB_2$ and the right hand sided peak represents $AlB_2$. It is concluded that for this concentration the sample is not purely single phase. The *c* parameter gets doubled and is 6.7157Å for the x = 0.40 sample and 6.7107Å for the x = 0.50 sample. The existence of this kind of doubled structure or superstructure was identified earlier by *HRTEM* (high resolution transmission electron microscopy) and *EDX* (Energy dispersive X-ray analysis) techniques [28-30]. Although the superstructure was also proposed by Serena Margadona et al. [31] based on a single super-lattice peak at a very low angle ($2\theta = 7.26^o$) but super-lattice peaks in the higher angle range were not observed/discussed. The only report about higher angle super structural peaks in $Mg_{0.5}Al_{0.5}B_2$ from *XRD* is by Xiang et al. [32]. Interestingly, Xiang et al., discussed only the $2\theta = 50^o$ to $55^o$ peak and marked the same within the $MgB_2$ structure, but did not consider the doubling of the *c* parameter or the possibility of a superstructure. Further, the observed shoulder of the so-called [002] peak was not considered. In fact a few extra lines at $2\theta = 62^o$ to $68^o$ were also seen in their spectrum but not discussed. Our *XRD* pattern for $Mg_{0.5}Al_{0.5}B_2$ cannot be indexed for all observed peaks without considering the double *c* parameter/superstructure. The simple model behind the doubling of the *c* parameter is as Mg-B-Al-B-Mg-B-Al-B. The hexagonal Mg layers are alternatively replaced in the lattice by the dimensionally (same *a* parameter) identical Al layers. Both crystallographically distinct $Mg^{2+}$ and $Al^{3+}$ ions present in the unit cell have the same co-ordination environments, namely, they lie directly above the centers of two boron hexagons of adjacent boron layers. Now the Mg-B-Al-B pattern is repeated in the lattice. The superstructure reflections are strongest for the crystals with compositions close to x = 0.5 in $Mg_{1-x}Al_xB_2$. In the specimen with overall composition x = 0.40, though the superstructure peaks are observed, the single cell reflections are seen as well. This shows that some portions of the x = 0.40 sample are Al rich and close to x = 0.50, exhibiting the superstructure peaks. On the other hand, eventually some portions have much less Al content than x = 0.40, resulting in a left side shifted [002] reflection of a single unit cell (see lower inset Fig. 1). Beyond x = 0.50, superstructure reflections fade away. No

superstructures are seen for the $Mg_{1-x}Al_xB_2$ system with 0.40 > x > 0.60. These results indicate that the appearance of superstructure is strongly constrained to a composition very close to $Mg_{0.5}Al_{0.5}B_2$.

The origin of the observed superstructure is of particular importance and may have consequences for the understanding of superconductivity in this system. Several causes can be proposed for the observed superstructure. The electronic instability originating due to continuous Al addition at 50% doping level might be one reason. This is consistent with the narrow stability range for the existence of superstructure and its high sensitivity to compositional changes. Secondly, it might be associated with a structural instability arising from the size mismatch between Al and Mg. Near x = 0.5 the size mismatch effect is maximum and hence the complete ordering of Mg and Al is energetically favorable.

Resistivity versus temperature plots of the $Mg_{1-x}Al_xB_2$ series of samples are shown in Fig. 4 for superconducting samples, i.e., with x = 0.0, 0.10, 0.20, and 0.40. Resistivity decreases with decreasing temperature and drops suddenly at their respective critical temperatures ($T_c$). In the normal state, i.e., above $T_c$, all samples show metallic behavior. The normal state $\rho$-$T$ plot of our pure $MgB_2$ and Al doped samples is in accordance with the published data on the system. [16-18]. Residual resistivity ratio (RRR = $\rho_{298}/\rho_{40}$) *values* for the $Mg_{1-x}Al_xB_2$ series are 2.58, 1.867, 1.71 and 1.42 for x = 0.0, 0.10, 0.20 and 0.40 samples, respectively. This clearly shows that RRR values decrease continuously with increasing Al content as shown earlier by Ref. [16]. The relatively low RRR value ~ 2- 3 as compared to values up to 20 reported for high purity $MgB_2$ and other ceramics seems to reveal a large contribution from impurity scattering [17]. The critical temperature $T_c$ ($\rho$ = 0) for the pristine sample is about 38 K. As we add aluminium, loss of superconductivity is observed in terms of decreasing critical temperatures. The exact variation of critical temperature with Al content (x in $Mg_{1-x}Al_xB_2$) is shown in the inset of Fig. 4(b). A continuous decrease in $T_c$ is observed which is in confirmation with the literature [15-18]. First, there is a slow decrease in $T_c$ up to x = 0.20, which is followed by a relatively sharper decrement untill x = 0.40. The samples are no longer superconducting beyond x = 0.40. The superconducting transition width is small for samples with x up to 0.20, but for the x = 0.40 sample, $T_c$ (onset) is 26.2 K while $T_c$ ($\rho$ = 0) is only 6.9 K. This decrease in $T_c$ and the exceptionally large transition width at x = 0.40 can readily be explained in terms of simple band filling due to electron doping by Al addition as reported earlier by [27,33-34]. No superconductivity is observed in the $Mg_{0.5}Al_{0.5}B_2$ sample.

Magnetization measurements carried out in both zero field cooled (ZFC) and field cooled (FC) situations for x= 0.0, 0.04, 0.10, 0.20, 0.25 and 0.40 in the $Mg_{1-x}Al_xB_2$ series are shown in Fig.6. In the zero field-cooled case, all samples show a diamagnetic transition at their critical temperatures as reported earlier [15, 28]. The critical temperatures obtained from here ($T_c^{dia}$) are in close agreement with the critical temperature ($T_c$ $\rho = 0$) obtained from resistivity measurements except in the case of x = 0.40 for which a broad transition is seen and saturation of the moment is not seen down to 5 K. Qualitatively, both $\rho$ - $T$ and $\chi$ - $T$ measurements confirm a decrease in $T_c$ with Al doping. Magnetization measurements done in the field-cooled situation differ from the zero field-cooled case due to trapped flux. The pure sample shows a very small paramagnetic signal at the same critical temperature. This is called the paramagnetic Meissner Effect and is explained elsewhere [35].

**Conclusion**

In summary, we report the structural and superconducting properties of Al substituted $MgB_2$ samples. Both resistivity and magnetization measurements show that Al substitution leads to the suppression of superconductivity in terms of a decrease in critical temperature. The structural analysis done by XRD revealed a continuous decrease in both lattice parameters and hence the cell volume, thus inducing a negative strain in the lattice. Moreover, Al substitution in the $MgB_2$ lattice resulted in a hexagonal superstructure at intermediate composition. This superstructure is very sensitive to the compositional changes and is constrained to x = 0.5 in the $Mg_{1-x}Al_xB_2$ series. The superstructure is analyzed in terms of ordering of Al and Mg layers. Size mismatch and electronic effects were considered as possible origins of the observed behavior.

**Acknowledgement**

The authors from *NPL* would like to thank Dr. Vikram Kumar (Director, *NPL*) for showing his keen interest in the present work. Authors would like to thank Dr. N.P. Lalla for his stimulating discussions regarding the identification of super-structural peaks at higher angle in XRD of our sample. One of us MM would also thank *CSIR* for financial support by providing *JRF* fellowship. Authors would also like to thank Prof. E. Takayama-Muromachi from *NIMS* Japan for helping in carrying out the *SQUID* magnetization measurements and S. Balamurugan for slow scan *XRD*.

**Figure and Table captions:**

Table 1: Lattice parameters table for $Mg_{1-x}Al_xB_2$ samples.

Table 2: Cell refinement report of Pure $MgB_2$ & $Mg_{0.5}Al_{0.5}B_2$ samples.

Fig.1 (a): X-ray diffraction pattern of $MgB_2$ showing all characteristic peaks.

Fig. 1(b): X-ray diffraction patterns of a series of $Mg_{1-x}Al_xB_2$ samples (x = 0.0 to 0.40). The two insets show partial shifts of (002) and (100) peaks. (The symbol # and * show some extra superstructural peaks and MgO peak respectively)

Fig.2: Change in lattice parameters with increasing Al content. Variation of c/a value is depicted in the inset.

Fig. 3: X-ray diffraction pattern of $Mg_{0.5}Al_{0.5}B_2$. Two insets are the magnified parts of the region that clearly shows superstructure peaks. (The symbol # and $ indicate the super structural peaks and unidentified peaks respectively.)

Fig. 4: Resistance versus Temperature curves for $Mg_{1-x}Al_xB_2$ samples (x = 0.0 to 0.40). The inset shows the variation of critical temperatures with Al content in $Mg_{1-x}Al_xB_2$ samples (x = 0.0 to 0.40).

Fig.5: Zero field cooled and Field cooled magnetization as a function of temperature for different Al doping level.

**Table 1:**

| Sample | *a* (Å) | *c* (Å) | *Volume* (Å $^3$) | *c/a* |
|---|---|---|---|---|
| $MgB_2$ | 3.0857(8) | 3.5280(8) | 29.09 | 1.143 |
| $Mg_{0.98}Al_{0.02}B_2$ | 3.0811(7) | 3.5175(7) | 28.92 | 1.142 |
| $Mg_{0.94}Al_{0.06}B_2$ | 3.0805(11) | 3.5153((11) | 28.89 | 1.141 |
| $Mg_{0.92}Al_{0.08}B_2$ | 3.0794(18) | 3.5043(18) | 28.78 | 1.137 |
| $Mg_{0.90}Al_{0.10}B_2$ | 3.0722(26) | 3.4977(25) | 28.59 | 1.138 |
| $Mg_{0.80}Al_{0.20}B_2$ | 3.0696(25) | 3.4653(24) | 28.28 | 1.128 |
| $Mg_{0.75}Al_{0.25}B_2$ | 3.0649(7) | 3.4338(6) | 27.93 | 1.120 |
| $Mg_{0.60}Al_{0.40}B_2$(+) | 3.055(17) | 3.446(16) | 27.86 | 1.128 |
| $Mg_{0.60}Al_{0.40}B_2$(*) | 3.062 (7) | 6.716(14) | 54.53 | *Sup-structure* |
| $Mg_{0.50}Al_{0.50}B_2$(*) | 3.052 (2) | 6.711(5) | 54.13 | *Sup-structure* |

+: Single cell concept, *: Double cell concept

**Table 2**

| **$MgB_2$, single cell Hexagonal $P_6$/mmm a = 3.088 Å, c = 3.528 Å** | **(hkl)** | **2θ(cal) (in degrees)** | **2θ(obs) (in degrees)** | **Delta** |
|---|---|---|---|---|
| | (001) | 25.223 | 25.172 | 0.051 |
| | (100) | 33.472 | 33.440 | 0.032 |
| | (101) | 42.370 | 42.356 | 0.014 |
| | (002) | 51.784 | 51.780 | 0.004 |
| | (110) | 59.837 | 59.836 | 0.001 |
| | (102) | 63.078 | 63.088 | -0.010 |
| | (111) | 65.976 | 65.984 | -0.008 |
| **$Mg_{0.5}Al_{0.5}B_2$, double cell Hexagonal $P_6$/mmm a = 3.052 Å, c = 6.711 Å** | | | | |
| | (001) | 13.183 | | |
| | (002) | 26.544 | 26.528 | 0.016 |
| | (100) | 33.886 | 33.864 | 0.022 |
| | (101) | 36.506 | 36.773 | -0.267 |
| | (003) | 40.286 | | |
| | (102) | 43.553 | 43.564 | -0.011 |
| | (103) | 53.631 | 53.764 | -0.133 |
| | (004) | 54.664 | 54.668 | -0.004 |
| | (110) | 60.630 | 60.636 | -0.006 |
| | (111) | 62.348 | 62.263 | 0.085 |
| | (104) | 65.888 | 65.824 | 0.064 |
| | (112) | 67.354 | 67.333 | 0.021 |

Cal- calculated
Obs- observed
Delta = 2θ cal - 2θ obs

Fig.1(a)

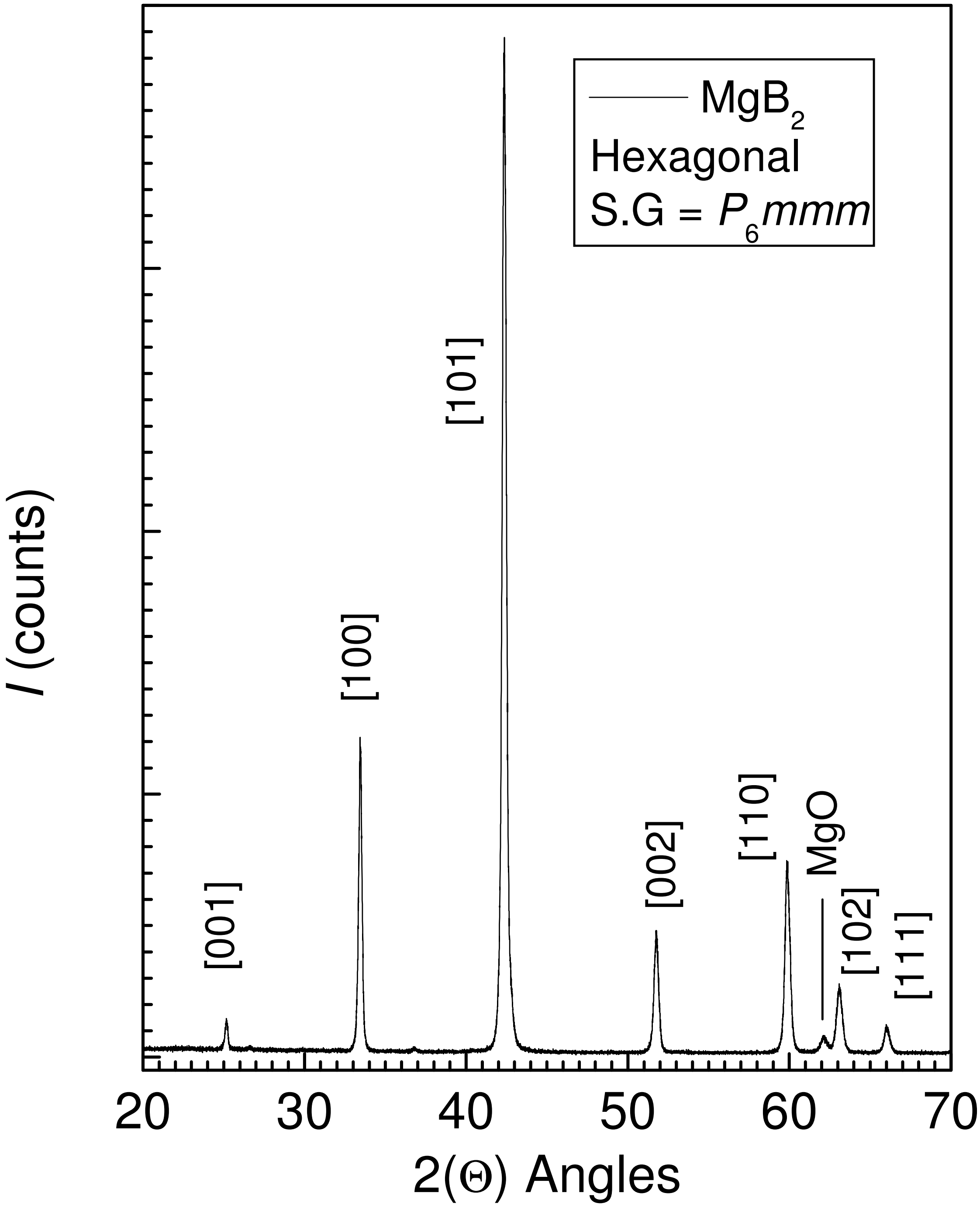

Fig. 1(b)

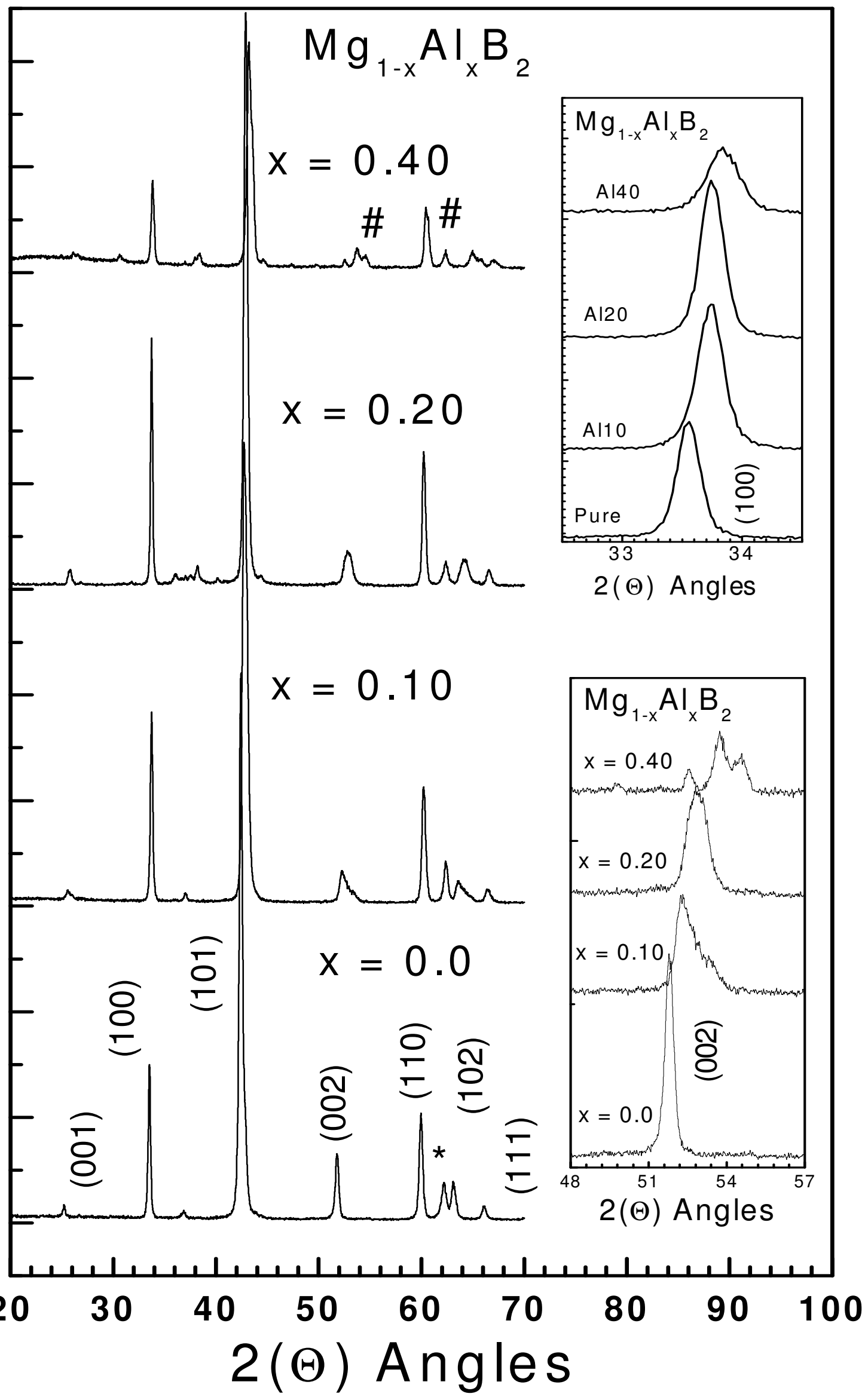

Fig.2

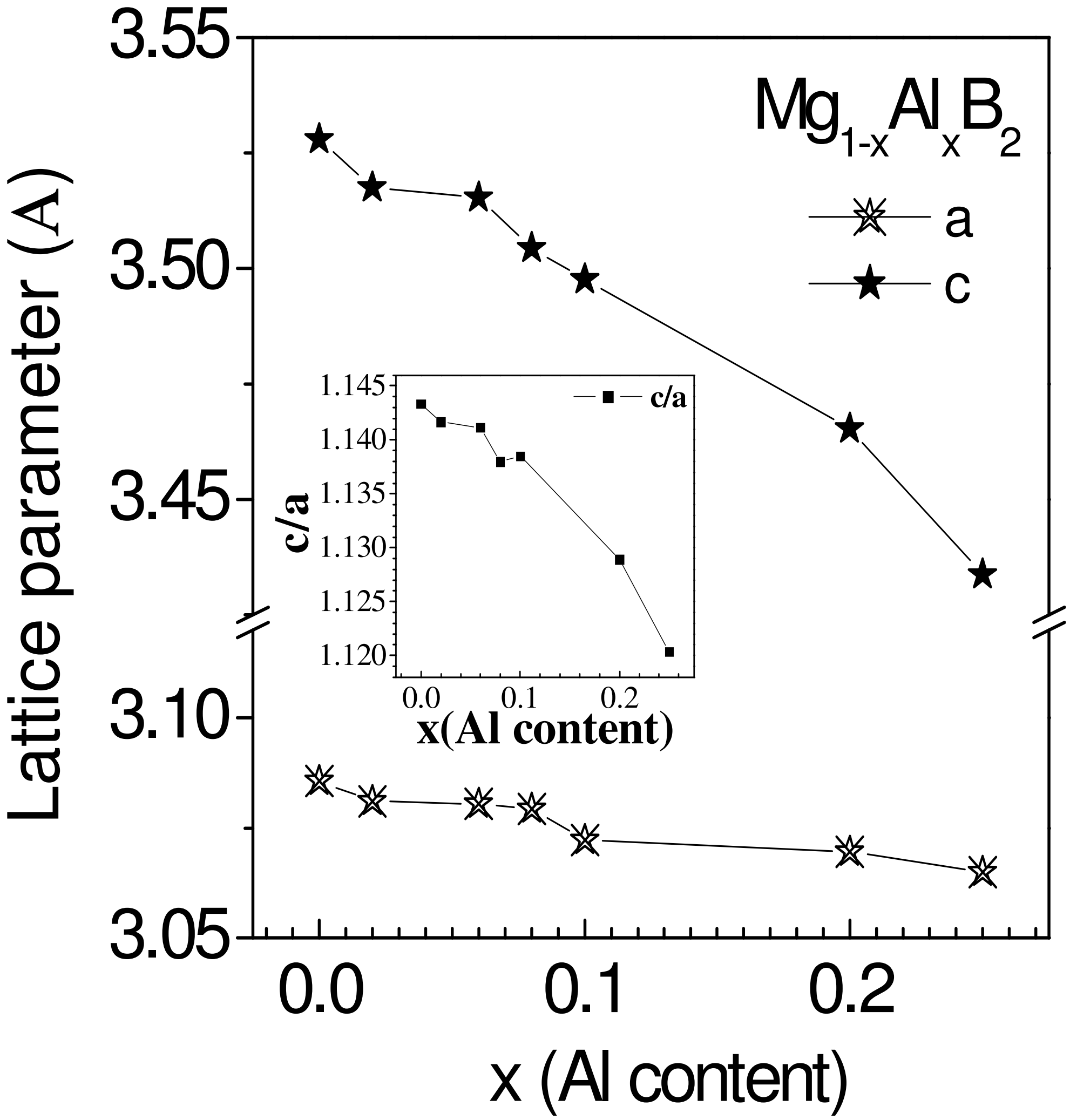

Fig. 3

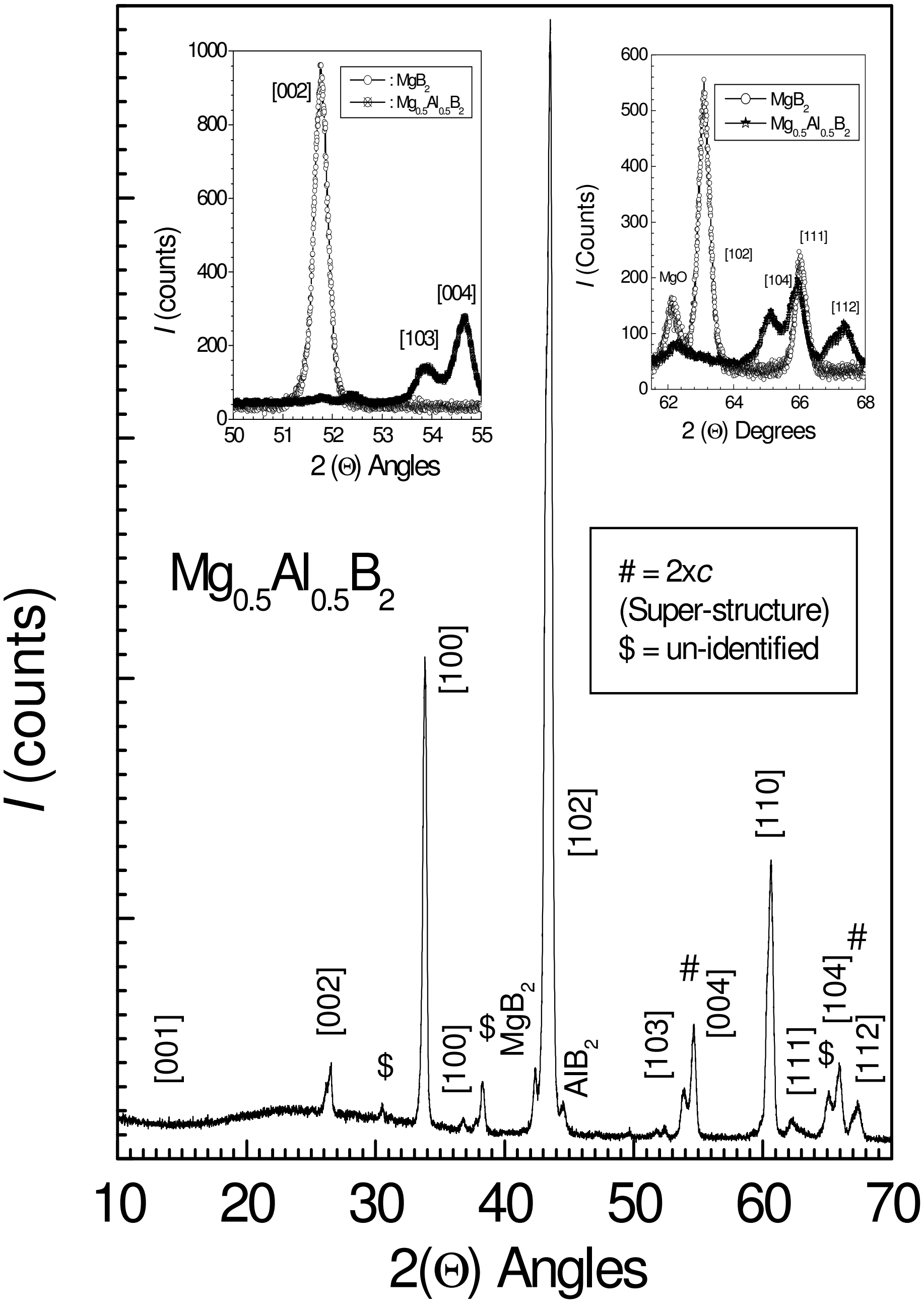

Fig.4

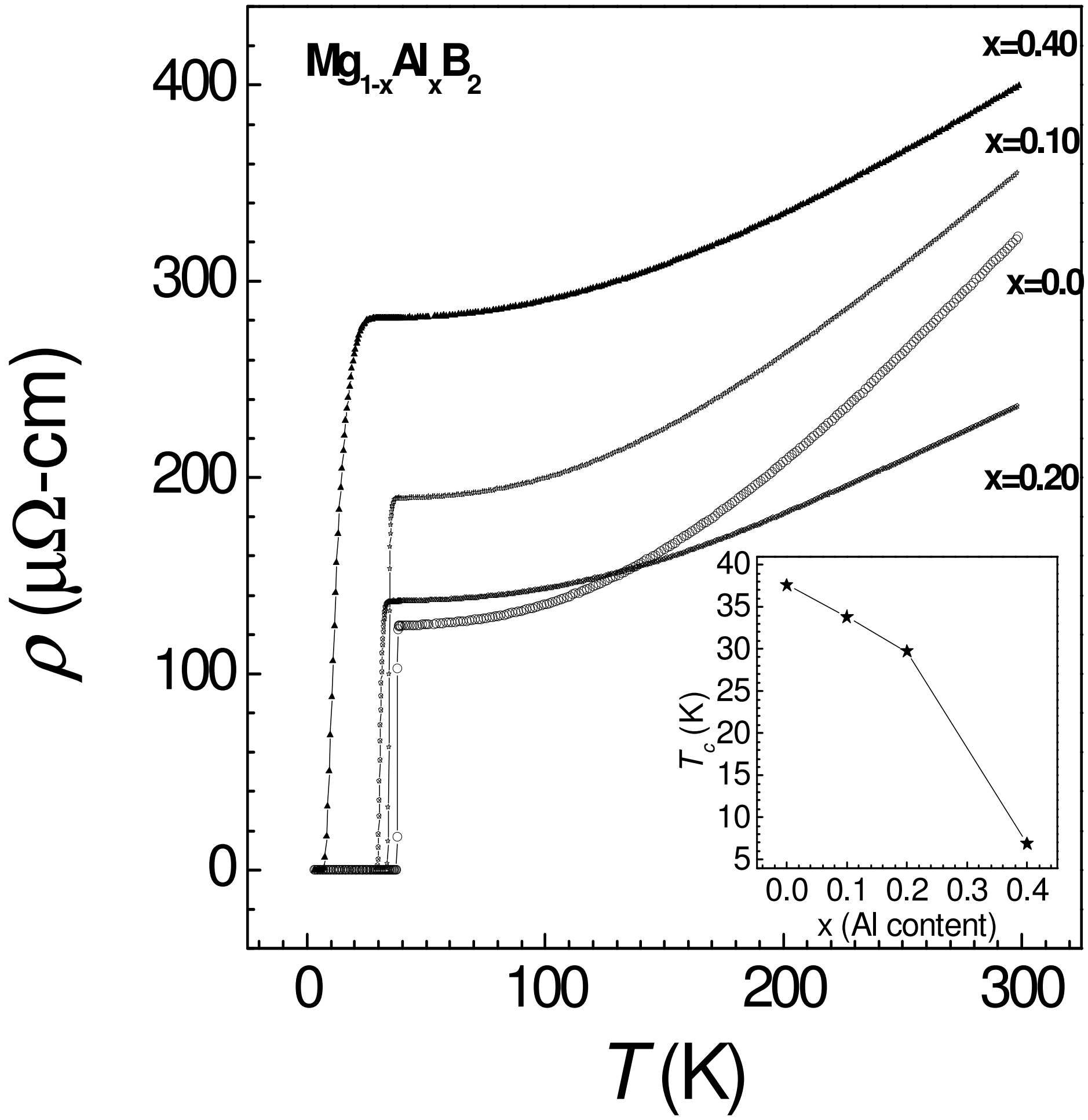

Fig. 5

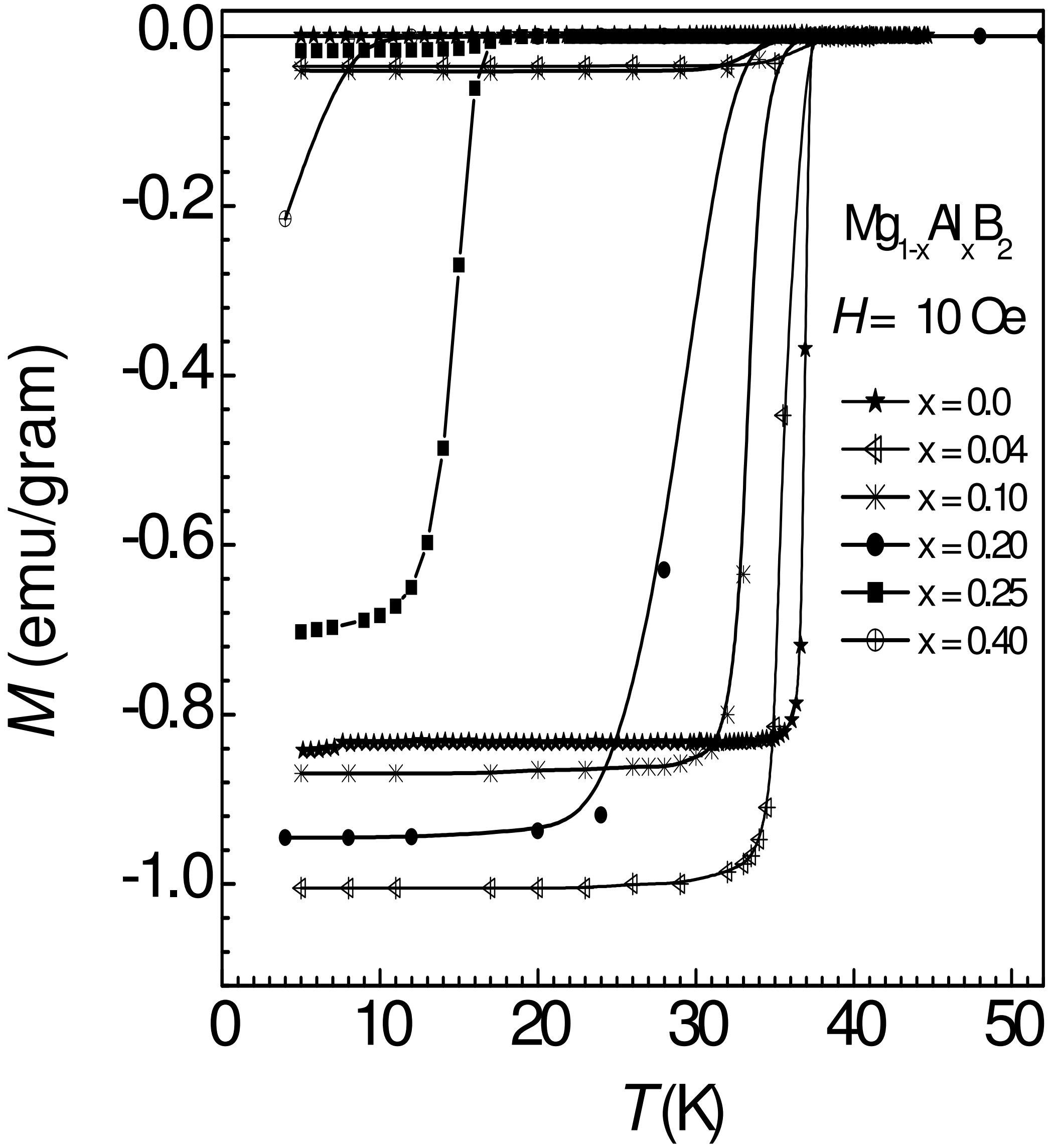